\documentclass[12pt]{article}
\usepackage{hyperref}
\usepackage[anythingbreaks]{breakurl}
\usepackage{color}
\usepackage{amsmath,amssymb,amsthm}
\usepackage{natbib}
\usepackage{array}
\usepackage{booktabs, multicol, multirow}
\usepackage[nohead]{geometry}
\usepackage[singlespacing]{setspace}
\usepackage[bottom]{footmisc}
\usepackage{floatrow}
\usepackage{float,graphicx}
\usepackage{caption}
\usepackage{indentfirst}

\newtheorem{theorem}{Theorem}

\newcommand{\beq}{\begin{equation}}
\newcommand{\eeq}{\end{equation}}

\title{Random problems with R}
\author{Kellie Ottoboni and Philip~B. Stark}
\date{\today}
\begin{document}
\maketitle

\begin{abstract}
\noindent R (Version 3.5.1 patched) has an issue with its random sampling functionality.
R generates random integers between $1$ and $m$
by multiplying random floats by $m$, taking the floor, and adding $1$ to the result.
Well-known quantization effects in this approach result in a 
non-uniform distribution on $\{ 1, \ldots, m\}$.
The difference, which depends on $m$, can be substantial.
Because the \texttt{sample} function in R relies on generating random integers,
random sampling in R is biased.
There is an easy fix: construct random integers directly from random bits, rather than
multiplying a random float by $m$.
That is the strategy taken in Python's \texttt{numpy.random.randint()} function, 
and recommended by the authors of the Mersenne Twister algorithm, among
others.
Example source code in Python is available at \url{https://github.com/statlab/cryptorandom/blob/master/cryptorandom/cryptorandom.py} (see functions \texttt{getrandbits()} and \texttt{randbelow\_from\_randbits()}).
\end{abstract}


A textbook way to generate a random integer on 
$\{1, \dots, m\}$ is to start with $X \sim U[0,1)$ and define $Y \equiv 1 + \lfloor mX \rfloor$. 
If $X$ is truly uniform on $[0,1)$, $Y$ is then uniform on $\{1, \dots, m\}$.
But if $X$ has a discrete distribution derived by scaling a pseudorandom $w$-bit integer
(typically $w=32$) or floating-point number, 
the resulting distribution is, in general, not uniformly distributed on 
$\{1, \ldots, m \}$ even if the underlying pseudorandom number generator 
(PRNG) is perfect.
Theorem~\ref{thm:theorem_1} illustrates the problem.

\begin{theorem}[\citet{knuth_art_1997}] 
\label{thm:theorem_1}
Suppose $X$ is uniformly distributed on $w$-bit binary fractions, and
let $Y_m \equiv 1 + \lfloor mX \rfloor$.
Let $p_+(m) = \max_{1 \le k \le m} \Pr\{Y_m = k\}$ and $p_-(m) = \min_{1 \le k \le m} \Pr\{Y_m = k\}$.
There exists $m < 2^w$ such that, to first order, 
$p_+(m)/p_-(m) = 1 + m2^{-w+1}$.
\end{theorem}

A better way to generate random elements of $\{1, \dots, m\}$ is to use pseudorandom bits directly,
avoiding floating-point representation, multiplication, and the floor operator. 
Integers between $0$ and $m-1$ can be represented with $\mu(m) \equiv \lceil \log_2(m-1) \rceil$ bits. 
To generate a pseudorandom integer between $1$ and $m$, first generate 
$\mu(m)$ pseudorandom bits (for instance, by taking the most significant $\mu(m)$ bits from the PRNG output, if $w \ge \mu(m)$, or by concatenating successive outputs of the PRNG and taking the
first $\mu(m)$ bits of the result, if $w < \mu(m)$).
Cast the result as a binary integer $M$.  
If $M > m-1$, discard it and draw another $\mu(m)$ bits; otherwise, return $M+1$.\footnote{%
   See \citet[p.114]{knuth_art_1997}.
   This is also the approach recommended by the authors of the Mersenne Twister. See
   \url{http://www.math.sci.hiroshima-u.ac.jp/~m-mat/MT/efaq.html}, last accessed 18~September 2018.
}
Unless $m = 2^{\mu(m)}$, this procedure is expected to discard some random draws---up to almost 
half the draws if $m = 2^p+1$ for some integer $p$.
But if the input bits are IID Bernoulli(1/2), the output will be uniformly distributed on $\{1, \ldots, m\}$.
This is how the Python function \texttt{numpy.random.randint()} (Version 1.14) generates pseudorandom integers.\footnote{%
  However, Python's built-in \texttt{random.choice()} (Versions 2.7 through 3.6) does 
  something else biased: it finds the closest integer to $mX$, where $X$ is a binary fraction 
  between 0 and 1.
}

The algorithm that R (Version 3.5.1 patched) \citep{R_2018} uses to generate random integers
in \texttt{R\_unif\_index()} (in \texttt{RNG.c})
has the issue pointed out in Theorem~\ref{thm:theorem_1} in a more complicated form, 
because R uses a pseudorandom float at an intermediate step, rather than multiplying a binary fraction
by $m$.
The way the float is constructed depends on $m$.
Because \texttt{sample} relies on random integers, it inherits the problem.

When $m$ is small, R uses \texttt{unif\_rand} to generate pseudorandom floating-point 
numbers $X$ on $[0, 1)$ starting from a $32$-bit random integer generated from the 
Mersenne Twister algorithm \citep{mt1998}.\footnote{ %
Luke Tierney pointed out that the seeding algorithm used in R is neither the one originally
proposed by \citet{mt1998}, which is known to have issues, nor their updated 2002 version
that fixes these issues.
Instead, R uses its own initialization method invented by Brian Ripley.} %
The range of \texttt{unif\_rand} contains 
(at most) $2^{32}$ values, which are approximately equi-spaced (but for the vagaries of converting
a binary number into a floating-point number~\citep{goldberg91}, which
R does using floating-point multiplication by 2.3283064365386963e-10).

When $m > 2^{31}$, \texttt{R\_unif\_index()}
calls \texttt{ru} instead of \texttt{unif\_rand}.\footnote{
   A different function, \texttt{sample2}, is called when $m > 10^7$ and $k < m/2$.
\texttt{sample2} uses the same method to generate pseudorandom integers.
}
\texttt{ru} combines two floating-point numbers, $R_1$ and $R_2$, each generated from a 32-bit integer, 
to produce the floating-point number $X$, as follows:
the first float is multiplied by $U = 2^{25}$, added to the second float, and the result is divided by
$U$:
$$ X = \frac{\lfloor U R_1 \rfloor + R_2}{U}.$$

The relevant code is in \texttt{RNG.c}.

The cardinality of the range of \texttt{ru} is certainly not larger than $2^{64}$.
The range of \texttt{ru} is unevenly spaced on $[0, 1)$
because of how floating-point representation works.
The inhomogeneity can make the probability that $X \in [x, x+\delta) \subset [0, 1)$
vary widely with $x$.

For the way \texttt{R\_unif\_index()} generates random integers, the non-uniformity of the probabilities of 
$\{1, \ldots, m\}$ is largest when $m$ is just below $2^{31}$. 
The upper bound on the ratio of selection probabilities approaches $2$ as $m$
approaches $2^{31}$, about 2~billion. 
For $m$ close to 1~million, the upper bound is about $1.0004$.

We recommend that the R developers replace the algorithm in \texttt{R\_unif\_index()} with the algorithm based on generating a 
random bit string large enough to represent $m$ and discarding integers that are larger than $m$.
The resulting code would be simpler and more accurate. 
Other routines that generate random integers using the multiply-and-floor method \texttt{(int) unif\_rand() * n}, 
for instance, \texttt{walker\_ProbSampleReplace()} in \texttt{random.c},
should also be updated to use an unbiased integer generator (e.g., to call the new 
version of \texttt{R\_unif\_index()}).

\bibliographystyle{plainnat}
\bibliography{refs}

\end{document}